\documentclass[a4paper,11pt]{article}
\usepackage{pos}
\usepackage{slashed}
\usepackage{enumitem}
\usepackage{natbib}

\title{Toward Compute-Bound Lattice-QCD Software}
\author*[a]{Anthony V.~Grebe}

\affiliation[a]{University of Maryland, College Park,\\
  4296 Stadium Dr, College Park, MD, USA}

\emailAdd{agrebe@umd.edu}

\abstract{Multi-right-hand-side multigrid inverters for lattice QCD are much less memory constrained than single-right-hand-side Krylov solvers, so the upper bound on efficiency of modern solvers is much higher than what was historically possible.  This work presents a reanalysis of roofline estimates to reflect this new reality.  It also describes the qc library, which implements multi-right-hand-side multigrid at an efficiency that can exceed 50\% of the processor roofline.}

\renewcommand{\hookAfterAbstract}{%
  \par\bigskip
  \textsc{Preprint}: UMD-PP-026-09
}

\FullConference{The 43rd International Symposium on Lattice Field Theory (Lattice 2026)\\
July 26 to August 1, 2026\\
University of Maryland, College Park, USA\\}

\begin{document}
\renewcommand{\logo}{\relax}
\maketitle

\section{Introduction}
Lattice QCD has for decades been one of the largest consumers of public supercomputing resources, with recent annual allocations reaching millions of node-hours on leadership-class machines.
Performance of lattice-QCD inverters, however, especially for Wilson-type fermions, has historically been about 10\% of the processor roofline for single-precision floating-point arithmetic.\footnote{
  Among CPU-based high-performance computers, Knights Landing processors were likely the most standardized, so many lattice-QCD packages were optimized for and benchmarked on such machines.  QPhiX performed $\slashed{D}$ applications at 435 GFlop/s on one 68-core KNL \cite{balint-portability}, about 8\% of the processor roofline of about 5 TFlop/s.  Bridge++ reports a maximum performance of 475 GFlop/s on 68-core KNLs (9\% of roofline) \cite{issaku}.  Domain-wall $\slashed{D}$ applications have performed better, with Grid's 960 GFlop/s achieving 16--18\% of the roofline \cite{grid, grid-performance}.  Performance of inverters is typically somewhat lower than $\slashed{D}$ application due to the additional dot products and vector additions needed.  GPU software is more complicated since only some parts of the calculation can be offloaded to very-high-speed low-precision tensor cores, making a single number for efficiency harder to define.
}
Analyses of memory traffic have indicated that this underutilization of the processor was an unavoidable consequence of the inherent structure of lattice-QCD software and that these performances were nearly optimal.
I will argue that these analyses were largely correct given the assumptions of the time but that such assumptions are no longer true for modern lattice-QCD algorithms.  As a result, substantially higher performances, reaching above 50\%, are now achievable.

\section{The Memory Roofline}
\label{sect:memory-roofline}
In order to perform arithmetic operations on lattice-valued fields that exceed the size of caches, processors must retrieve inputs from and store outputs to RAM.  The ratio of single-precision flops that can be performed per byte fetched from RAM is a measure of how memory-bandwidth-constrained a system is, and this ratio is about 10--30 flops per byte (F/B) for both CPU and GPU machines and in both enterprise-grade and consumer-grade hardware.

The flop-per-byte ratio that the hardware can provide must be compared to the arithmetic intensity (AI) of the algorithm, the flop-to-byte ratio that software demands, to determine whether code is memory-bound or compute-bound.  
Vector addition in single precision consumes 1 flop for 12 bytes of memory traffic (for a solidly memory-bound AI of 0.083 F/B), whereas complex matrix multiplication of $N\times N$ matrices consumes about $8N^3$ flops while reading $24N^2$ bytes, for an AI of $N/3$ F/B, which is compute bound for large $N$ ($N \gtrsim 60$ on most hardware).

In lattice QCD, the $\slashed{D}$ stencil in the Wilson-clover action is given by
\begin{equation}
  \slashed{D} \psi(x) = -\kappa \sum_{\mu = \pm 1}^{\pm 4} U_\mu (x) (1 - \gamma_\mu) \psi(x+\hat\mu) + F(x) \psi(x) \, ,
  \label{eq:Dslash}
\end{equation}
where $F(x) = 1 + \frac{1}{2}\kappa c_\text{SW} \sigma_{\mu\nu} F_{\mu\nu}(x)$.
$(1\pm\gamma_\mu)/2$ is a spin projection matrix: it compresses the information in 4 spins down to 2, so the gauge link is only applied to a 2-component spinor rather than a 4-component one.  The flop count is then 12 flops for the projection (6 complex additions), 132 flops for the gauge link application (a $3\times 3$ complex matrix times a $3\times 2$ one), and 24 flops for the reconstruction and accumulation into the output, for 168 flops per direction, plus 24 for multiplying the final solution by $\kappa$, giving 1368 total.  The clover term $F(x)$ can be stored as a pair of $6\times 6$ Hermitian matrices that are applied to the upper and lower 2-spinors of $\psi(x)$, costing $276\times 2 = 552$ flops, for a total of 1920 flops per site for the full Wilson-clover operator.\footnote{
  References in the literature disagree on this flop count by about 5\%.  The numbers here match the conventions used by QUDA \cite{quda} and DD$\alpha$AMG \cite{ddalphaamg-1}.  In particular, the matrix multiplications here do not count flops for accumulation into an output (Ref.~\cite{issaku} does include this and counts 1944 flops/site).  On the other hand, this does count the multiplication by $\kappa$, which could be avoided by folding into gauge links.  Also, the Hermitian matrices in $F(x)$ have a purely real diagonal, but the flop count above assumes they are general complex matrices.

  The flop count of the Wilson hopping term is reduced in a basis in which one of the spin projectors is diagonal (e.g.~the time direction in the Dirac basis), but then a basis conversion is needed for the clover term, costing the saved flops.  More drastically, axial gauge fixing eliminates the gauge multiplications in all but one of the links in one direction (typically the temporal one) \cite{quda}, reducing the flop count by 264.  This is left open as an avenue for optimization in future work.

  Even-odd preconditioning does not appreciably change the flop count per $\slashed D$ application --- this halves the number of sites, but per remaining site, the even-site operator does approximately double the amount of computational work.
}

The memory traffic na\"ively requires reading nine sites ($x$ and its eight neighbors), the eight $3\times 3$ gauge matrices,\footnote{
  Generic $3\times 3$ matrices are 18 reals, but gauge links in $SU(3)$ can be compressed to 12 reals (the first two rows) or even 8 (generators of the algebra) and reconstructed on the fly, saving memory at the cost of extra flops.  This is a net win for applications of $\slashed D$ to a single $\psi$ on memory-bound machines.  For multi-RHS solvers (see below), however, gauge reads become subdominant traffic, negating the benefit of such a scheme.
}
the clover term (two $6\times 6$ Hermitian matrices) and write one site as output, for a total of 456 real numbers, or 1824 bytes in single precision.  However, the inputs are reused across sites, so under perfect conditions the amortized traffic per site is a single-site fermion field for each of input and output, four gauge links, and the clover term, for a total of 768 bytes per site.  Depending on cache reuse, the arithmetic intensity of applying $\slashed D$ is then 1.05--2.5 flops per byte \cite{issaku}, about 5--20\% of what the memory can supply.\footnote{
  Perfect cache reuse is especially difficult in four-dimensional problems like lattice QCD with high surface-to-volume ratios of the working set.  Empirically, I have found that reading a fermion during $\slashed D$ applications costs about the same as 2--3 reads of the input field, compared to 1 read under perfect conditions and 9 under pessimal ones.
}

Multi-right-hand-side (multi-RHS) solvers that simultaneously apply $\slashed D$ to multiple fermion fields $\psi$ amortize the reads of the gauge and clover fields to improve the arithmetic intensity \cite{multi-rhs-1, multi-rhs-2, multi-rhs-3}.  Ignoring cache reuse, fermion reads dominate memory traffic, so multi-RHS gains at most a factor of 2, cutting bytes read from 1824 to $960 + 864/N$ amortized over $N$ RHS, consistent with Ref.~\cite{ddalphaamg-2026}.  Perfect cache reuse allows larger gains, from 768 bytes to $192 + 576/N$, a fourfold reduction for large $N$.  The optimal scenario -- many right-hand sides with perfect cache reuse -- gives a maximum arithmetic intensity of 10 flops/byte, comparable to hardware limitations.  Even a more realistic assumption of 5 flops/byte due to imperfect caching and finite $N$ would allow 15--50\% efficiency.

\section{Multigrid}
A major advance in lattice QCD in recent years has been the introduction of multigrid methods to reduce the time to solution, especially at light quark masses \cite{brannick-mg, babich-mg, ddalphaamg-1, ddalphaamg-2, ddalphaamg-thesis, quda-mg}.\footnote{Multigrid methods are beneficial even at relatively heavy quark masses; my own tests have found QUDA's multigrid solver \cite{quda-mg} to outperform BiCGStab by a factor of about 2 at strange-quark masses on coarse ensembles.}
The cost of standard Krylov solvers (CG, BiCGStab, etc.) is dominated by low (IR) modes of the theory from light pions travelling long distances, which can be removed by a coarser grid that captures this long-distance behavior.  A separate so-called ``smoother'' acts on the fine grid to remove the short-distance (UV) modes.  Thus, the standard multigrid algorithm is
\begin{enumerate}[noitemsep]
  \item Apply the restrictor $R$ to a fine-grid fermion to generate a coarse-grid approximation
  \item Approximately invert the coarse-grid operator $D_C = R D P$ against this fermion
  \item Apply the prolongator $P = R^\dagger$ to reconstruct the fine-grid solution
  \item Apply a post-cycle smoother on the fine grid to dampen UV modes
\end{enumerate}
In order to capture the IR modes of the theory, $R$ is typically built from $M$ near-null vectors.  Computing the dot product of each of these near-null vectors against the chirality-projected fermion (i.e.~either the upper or lower spin components in a basis where $\gamma_5$ is diagonal) on each aggregate (fine-grid block) replaces the 12 spin-color components at each fine-grid site with $2M$ components at each coarse grid site.  For $M = 32$ and fine-grid block sizes of $4^4$, this represents about a 50-fold compression of the fermion size (and therefore in the amount of memory traffic from fermions).

On medium to large lattice sizes, the two-level multigrid described above can be improved by performing the approximate inversion in step 2 above with an inner multigrid hierarchy, typically with a block size of about $2^4$.  This can be recursed as needed until the innermost level is small enough that additional levels are no longer beneficial.

Since the multigrid hierarchy has already damped the IR modes of the theory, the post-smoother only needs to remove the UV modes.  Applying the full $\slashed D$ operator to the fermion is sufficient to do this but not necessary.  An alternative that is more focused on UV modes is domain decomposition via the Schwarz alternating procedure (SAP) \cite{luscher-sap, balint-sap}: subdividing the lattice into smaller blocks, preferably those used in multigrid, and (approximately) solving within each block with its boundary clamped to zero.  SAP was coupled to inexact deflation, an early two-level method, in Ref.~\cite{luscher-deflation} and later to multigrid solves in the lattice-QCD library DD$\alpha$AMG \cite{ddalphaamg-1, ddalphaamg-2, ddalphaamg-thesis}.

\subsection{Arithmetic Intensity of Multigrid}
The coarse operator $D_C$ is a $2M \times 2M$ matrix at each coarse site and each coarse link.  With a single RHS, this is effectively a matrix-vector multiplication dominated by memory traffic of the matrix, reading $8M^2$ real elements of the matrix and performing $4M^2$ complex multiplications or $32M^2$ real flops, for an arithmetic intensity of about 1.  With optimal cache reuse, each link matrix can be used twice (once per endpoint), increasing this to 2, but well below the processor roofline.

With multiple RHS, even ignoring cache reuse, $D_C$ applications become dense matrix multiplications of a $2M \times 2M$ operator matrix by a $2M \times N$ fermion matrix, for $4M(2M+N)$ total reads (plus well-amortized output traffic).  This gives an arithmetic intensity of $32 M^2 N / 16M(2M+N) = 2MN/(2M+N)$ flops per byte, reaching 12.8 flops/byte at $M=32, N=16$ and asymptoting to $2M$ as $N\rightarrow \infty$.  The arithmetic is similar for the prolongation and restriction kernels.  The coarse multigrid levels can therefore show order-of-magnitude speedups on memory-bound systems with multi-RHS solvers, substantially exceeding the benefits seen by fine-grid Krylov solvers \cite{boyle2024multiplerighthandmultigrid, kate-lattice-2024}.

On the fine grid, the arithmetic intensity can be substantially boosted by using the domain-decomposed Schwarz alternating procedure as the post-smoother, as the per-block solves (typically done iteratively, at least on the fine grid) are cache local.  
The first iteration streams the fermion and gauge fields from memory, but on a $4^4$ block with 16 RHS, the fields total $\sim 1$ MB and mostly reside in L2 or L3 cache, increasing the arithmetic intensity of $\slashed D$ by a factor of the number of inner SAP iterations, typically 3--4, so all iterations after the first are not memory bound and can run at a speed comparable to the processor roofline.

\section{The qc Library for Lattice QCD}
DD$\alpha$AMG was written about a decade ago as an implementation of the domain-decomposed multigrid algorithms in Refs.~\cite{ddalphaamg-1, ddalphaamg-2, ddalphaamg-thesis}.  It was designed to support a single RHS at a time (as was the norm in lattice-QCD solvers at the time) and also targeted pre-AVX machines, so vectorization was limited to SSE.

qc\footnote{Named after Qlua \cite{qlua} but in C.}
is designed as a more performant multi-RHS reimplementation of DD$\alpha$AMG on modern hardware.
As in QUDA \cite{quda, kate-lattice-2024}, Krylov solvers use different coefficients for each RHS throughout, and the stopping criterion is based on the slowest of the solves.
Modern machines require a vector width of 8 (for AVX2) or 16 (for AVX-512) to maximize single-precision performance, and empirically, I have found performance is best when vectorizing over the RHS rather than, e.g., the source position.  This avoids cross-lane operations for fermion shifts in the direction of vectorization, but it does require that the number of RHS be a multiple of the vector width.  A single source with 12 spin-color components does not lend itself naturally to 16 RHS, but since most lattice-QCD calculations (especially those using multigrid, where setup time is nontrivial) perform multiple solves per configuration, buffering sources appropriately can surmount this restriction.

During the fine-grid $\slashed D$ applications, qc performs a tiled traversal of the lattice to improve cache reuse, and it prefetches the next direction at each site to hide latency.  The inner kernels are written in assembly for AVX2 and AVX-512 architectures, reflecting the observation in Grid that assembly performs better than C intrinsics \cite{grid, grid-performance}, with plain C as a fallback for generic targets.
Implementation files are compiled in both single- and double-precision translation units, mirroring the design of Qlua \cite{qlua}.  The geometry, RHS count, and multigrid structure are \verb|#define|d at compile time so that compilers can optimize the computation of memory addresses for site indices.  Recompilation time is deliberately kept low (about two minutes) to facilitate changing ensembles or retuning.

A significant algorithmic change from DD$\alpha$AMG is the use of exact dense inverses (with even-odd preconditioning within each block) to replace the iterative solver in coarse-grid SAP.  On a $2^4$ block, the dense inverse performs about as much work as two iterations of the minimal-residual (MR) solver within each block (compared to about 3--4 in iterative approaches), and it provides an exact solution up to floating-point roundoff.  In a similar vein, after recursing down to a small bottom solve ($2^3 \times 4$ or similar size), the iterative bottom solver can also be replaced by a dense inverse.  Since the cost of such an inverse scales quadratically in block size, these are ill suited to a $4^4$ block solve during fine-grid SAP, but they can save significant computational work (and improve accuracy) during the smaller SAP blocks used on coarser levels.\footnote{
  The cubic cost of performing the inversion is paid only during setup, so it is amortized over all solves on the configuration and, like the rest of setup cost, is excluded from the performance analysis.
}

The qc library is released on GitHub at \url{https://github.com/agrebe/qc} under a GNU General Public License.  This publication corresponds to software version 1.0.0.

\subsection{Performance}
qc was tested on the USQCD CPU machines \texttt{LQ1} at Fermilab (based on the Cascade Lake architecture) and \texttt{lqcd} at BNL (Zen 4) as well as \texttt{Perlmutter} at NERSC (Zen 3).  Properties of these machines are listed in Tab.~\ref{tab:machines}.

\begin{table}
  \centering
  \begin{tabular}{|c|c|c|c|} \hline
    Machine & \texttt{LQ1} & \texttt{lqcd} & \texttt{Perlmutter} \\ \hline
    Location & FNAL & BNL & NERSC \\ \hline
    Architecture & Cascade Lake & Zen 4 & Zen 3 \\ \hline
    Cores/Node & 40 & 64 & 128 \\ \hline
    Frequency (GHz) & 2.5 & 3.75 & 2.67 \\ \hline
    Max fp32 Flops/Cycle & 64 & 32 & 32 \\ \hline
    Vectorization & AVX-512 & AVX-512 & AVX2 \\ \hline
    Processor Roofline (fp32 TFlop/s) & 6.4 & 7.7 & 10.9 \\ \hline
%    NUMA Domains/Node & 2 & 8 & 8 \\ \hline
%    RAM (GB) & 192 & 1536 & 512 \\ \hline
    STREAM Memory Bandwidth (GB/s) & 200 & 761 & 332 \\ \hline
    Flops/Byte Achievable & 32 & 10.1 & 32.9 \\ \hline
  \end{tabular}
  \caption{Machines used for testing qc.  Memory speeds are measured with STREAM Triad \cite{stream} with one copy of the executable per NUMA domain, and frequencies are measured under all-core FMA load.}
  \label{tab:machines}
\end{table}

Two test configurations with nearly physical valence quark masses ($m_\pi \approx 170$ MeV) were used in benchmarking: (A) a smeared $32^3 \times 48$ clover-on-HISQ configuration generated by the MILC collaboration \cite{milc} with $\kappa = 0.12575$, $c_\text{SW} = 1.0$ as in Ref.~\cite{npipi} and (B) a $64^3 \times 128$ isoclover ensemble generated by NME with parameters in Ref.~\cite{isoclover}.  Run parameters are given in Tab.~\ref{tab:configs}.  

\begin{table}
  \centering
  \begin{tabular}{|cc|c|c|} \hline
    Label & & A & B \\ \hline
    $(L/a)^3 \times (T/a)$ & & $32^3 \times 48$ & $64^3 \times 128$ \\ \hline
    $\approx a$ (fm) & & 0.15 & 0.09 \\ \hline
%    $am_q$ & & $-0.023856859$ & $-0.245$ \\ \hline
%    $c_\text{sw}$ & & 1.0 & 1.20536588031793 \\ \hline
    \# Levels & & 4 & 5 \\ \hline
    MG blocking & & $4^4$ / $2^4$ / $2^3 \times 3$ & $4^4$ / $2^4$ / $2^4$ / $2^4$ \\ \hline
    Null vectors & & 48 / 96 / 256 & 48 / 80 / 128 / 256 \\ \hline
    Nodes & $ \left\{ \begin{matrix}\texttt{LQ1} \\ \texttt{lqcd} \\ \texttt{Perlmutter} \end{matrix} \right. $ & $ \begin{matrix} 1/2 \\ 1/8 \\ 1/8 \end{matrix} $ & $ \begin{matrix} 8 \\ 1 \\ 4 \end{matrix} $ \\ \hline
    Outer Iterations & & 10 & 13 \\ \hline
    Time Used (node-s) & $ \left\{ \begin{matrix}\texttt{LQ1} \\ \texttt{lqcd} \\ \texttt{Perlmutter} \end{matrix} \right. $ & $ \begin{matrix} 7.81 \\ 2.95 \\ 3.85 \end{matrix} $ & $ \begin{matrix} 280 \\ 92.8 \\ 166 \end{matrix} $ \\ \hline
    Performance (TFlop/s/node) & $ \left\{ \begin{matrix}\texttt{LQ1} \\ \texttt{lqcd} \\ \texttt{Perlmutter} \end{matrix} \right. $ & $ \begin{matrix} 1.54~(24\%)\\ 4.09~(53\%) \\ 3.13~(29\%) \end{matrix} $ & $ \begin{matrix} 1.18~(18\%)\\ 3.58~(47\%) \\ 2.00~(18\%)\end{matrix} $ \\ \hline
  \end{tabular}
  \caption{A comparison of performance of multigrid solves using the qc library on two nearly-physical-mass configurations.  
  The fractional node counts for configuration A reflect running multiple copies of the executable (one per NUMA domain).
  The parameters above were found to maximize speed of convergence of the solver on \texttt{lqcd} (excluding time for setup and full-solution reconstruction from even-odd preconditioned fermions) over a reasonable search.  On the other machines, parameters match \texttt{lqcd} except that the bottom solve on configuration B was increased in size on \texttt{LQ1} to allow parallelism over sufficient MPI ranks (\texttt{Perlmutter} avoids this by running fewer MPI ranks per node on B).  Flops are counted exactly for coarse-grid and transfer operations and use the nominal values in Sect.~\ref{sect:memory-roofline} for fine-grid $\slashed D$, and the efficiencies are computed relative to the machine's single-precision processor roofline in Tab.~\ref{tab:machines}.
All solves were performed to a precision of $\varepsilon = 10^{-10}$.
}
  \label{tab:configs}
\end{table}

Performance for configuration A on a single rank was 24\% on \texttt{LQ1}, 53\% on \texttt{lqcd}, and 29\% on \texttt{Perlmutter}.  Performances fell about 6 percentage points on \texttt{LQ1} and \texttt{lqcd} and about 11 percentage points on \texttt{Perlmutter} when moving to configuration B on multiple MPI ranks.\footnote{On \texttt{Perlmutter}, the small bottom solve of configuration B limited the number of MPI ranks and forced same-socket NUMA domains to share an MPI rank, which likely led to the worse scaling.  Remedies for this, including the ability to divide the lattice in the $y$-direction rather than just $z$ and $t$, are being explored in ongoing work.}  On memory-constrained machines (\texttt{LQ1} and \texttt{Perlmutter}), the performance corresponds to about 8--9 flops/byte on configuration A.  On \texttt{lqcd}, with higher memory bandwidth,\footnote{
  It is also worth noting here that \texttt{lqcd}'s configuration was specified by Peter Boyle for lattice-QCD workloads, giving it a higher-than-average memory bandwidth and therefore enabling high performance on that machine.
}
this corresponds to about 5 flops/byte, showing that performance is moving into the compute-bound (rather than memory-bound) regime.

\section{Summary and Ongoing Work}
This work presents the qc library as a constructive proof that modern lattice-QCD algorithms are substantially less memory bottlenecked than traditional single-RHS Krylov solvers and can therefore achieve performances that are a sizeable fraction of processor rooflines.  Given the large expenditure on computing and the commensurately large carbon footprint of lattice-QCD calculations, such efficiency gains can translate into real environmental benefits.\footnote{A million-node-hour allocation on Frontier consumes about 2.6 GWh of electricity (based on a full-machine draw of 24.6 MW \cite{top500}), equivalent to about 940 metric tons of CO$_2$ based on US averages \cite{eia}.}

Further work is ongoing to explore additional performance optimizations.  Moving to lower precision in the inner solver has been shown to accelerate mixed-precision solvers \cite{issaku-half, quda} without sacrificing outer-solver accuracy.  This is especially important on GPUs, with low-precision tensor cores 1--2 orders of magnitude faster than single-precision arithmetic.  Work is therefore underway to support lower-precision operations and to port qc to GPUs.

\acknowledgments
I am deeply grateful to all the lattice-QCD software developers who have come before me.  In particular, Bridge++ \cite{issaku}, Chroma \cite{chroma}, Grid \cite{grid}, Qlua \cite{qlua}, QPhiX \cite{qphix}, QUDA \cite{quda}, and especially DD$\alpha$AMG \cite{ddalphaamg-1, ddalphaamg-2} informed many of the design decisions of qc and were essential in benchmarking and validation.  I would also like to thank Peter Boyle, Kate Clark, Will Detmold, Dan Hackett, B{\'a}lint Jo{\'o}, Gurtej Kanwar, Issaku Kanamori, Jo Moscoso, Robert Perry, Andrew Pochinsky, and Michael Wagman for useful conversations.  This work was made possible by computational resources from USQCD (especially Fermilab and BNL) and the resources of the National Energy Research Scientific Computing Center (NERSC), a Department of Energy Office of Science User Facility using NERSC award NP-ERCAPm747, and funding from the National Science Foundation under Grant No.~PHY-2402275.

Claude Code was used for code implementation and in the review of this manuscript.

\bibliographystyle{JHEP}
\bibliography{refs}

\end{document}